\documentclass[9pt,conference,a4paper]{IEEEtran}
\ifCLASSINFOpdf
   \usepackage[pdftex]{graphicx}
\else
\fi
\usepackage[tight,footnotesize]{subfigure}

\usepackage{amssymb}
\usepackage{amsmath}
\usepackage{url}

\usepackage{cite}
\usepackage{color}

\begin{document}

\title{Analysing the polarisation of the CMB \\with spin scale-discretised wavelets}

\author{%
\IEEEauthorblockN{
Boris Leistedt\IEEEauthorrefmark{1},  
Jason D. McEwen\IEEEauthorrefmark{2}, 
Martin B\"{u}ttner\IEEEauthorrefmark{1}, 
Hiranya V. Peiris\IEEEauthorrefmark{1}, 
Pierre Vandergheynst\IEEEauthorrefmark{3},
and Yves Wiaux\IEEEauthorrefmark{4}.
}
\\
\IEEEauthorblockA{\IEEEauthorrefmark{1} Department of Physics and Astronomy, University College London (UCL), London WC1E 6BT, UK}
\IEEEauthorblockA{\IEEEauthorrefmark{2} Mullard Space Science Laboratory (MSSL), University College London (UCL), Surrey RH5 6NT, UK}
\IEEEauthorblockA{\IEEEauthorrefmark{3} Institute of Electrical Engineering, Ecole Polytechnique F\'{e}d\'{e}rale de Lausanne (EPFL), CH-1015 Lausanne, Switzerland}
\IEEEauthorblockA{\IEEEauthorrefmark{4} Institute of Sensors, Signals \& Systems, Heriot-Watt University, Edinburgh EH14 4AS, UK}
}

\maketitle

\begin{abstract}
We discuss a new scale-discretised directional wavelet transform to analyse spin signals defined on the sphere, in particular the polarisation of  the cosmic microwave background (CMB).
\end{abstract}

\section{Spin scale-discretised wavelets on the sphere}

\begin{figure}[h!]
\centering
\subfigure[Spin-0 scale-discretised (directional) wavelets]{\hspace*{-2mm}\includegraphics[width=9.0cm]{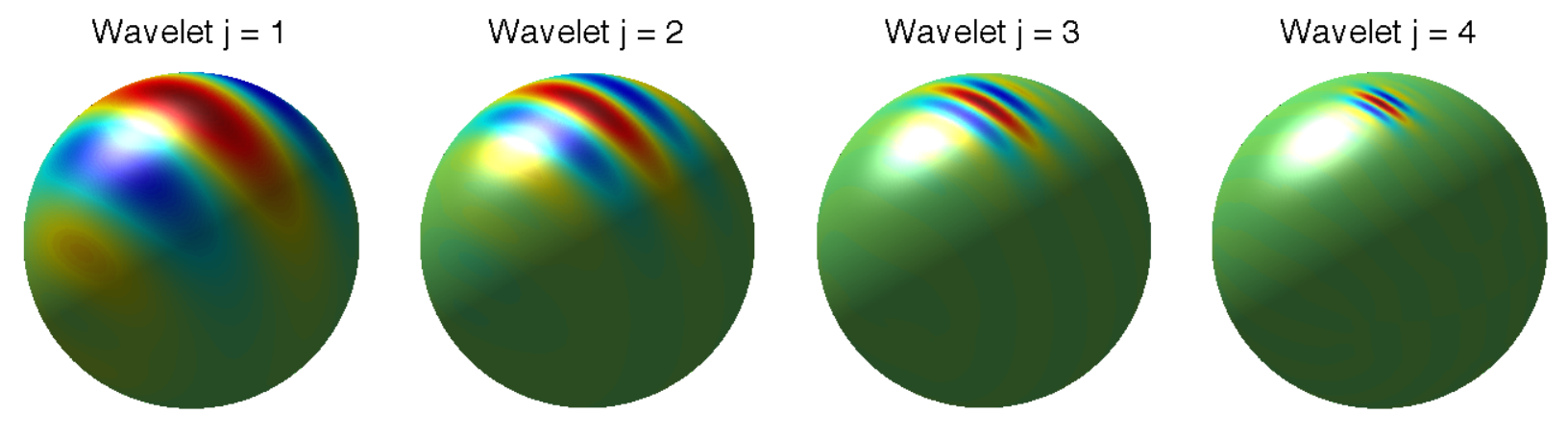}}\vspace*{2mm}
\vspace*{1mm}
\subfigure[Spin-2 scale-discretised (directional) wavelets]{\hspace*{-2mm}\begin{minipage}{9.0cm}\includegraphics[width=9.0cm]{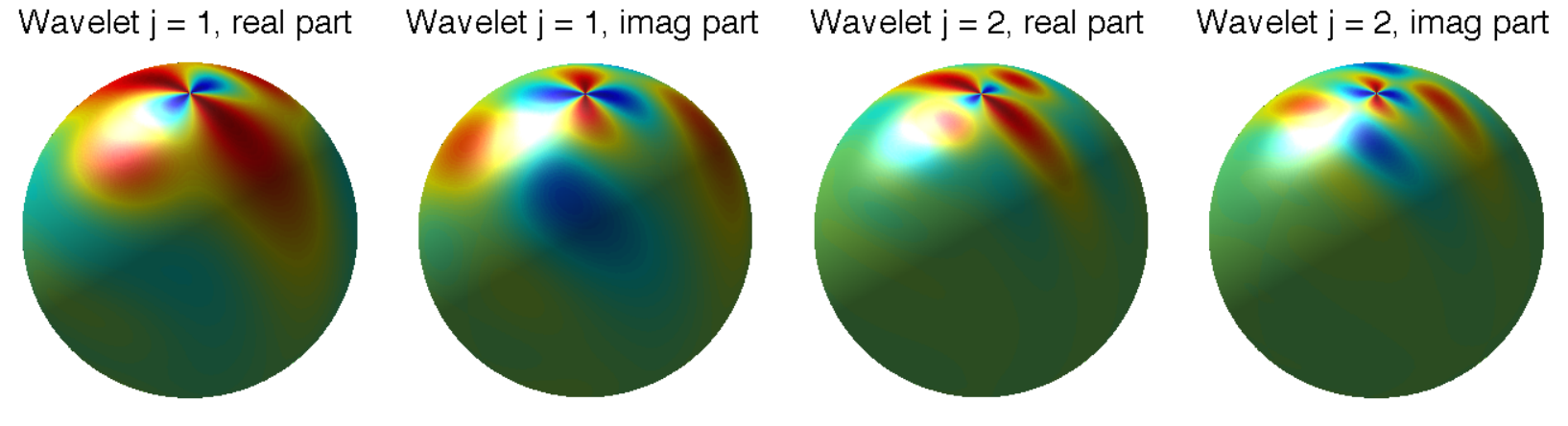}
\includegraphics[width=9.0cm]{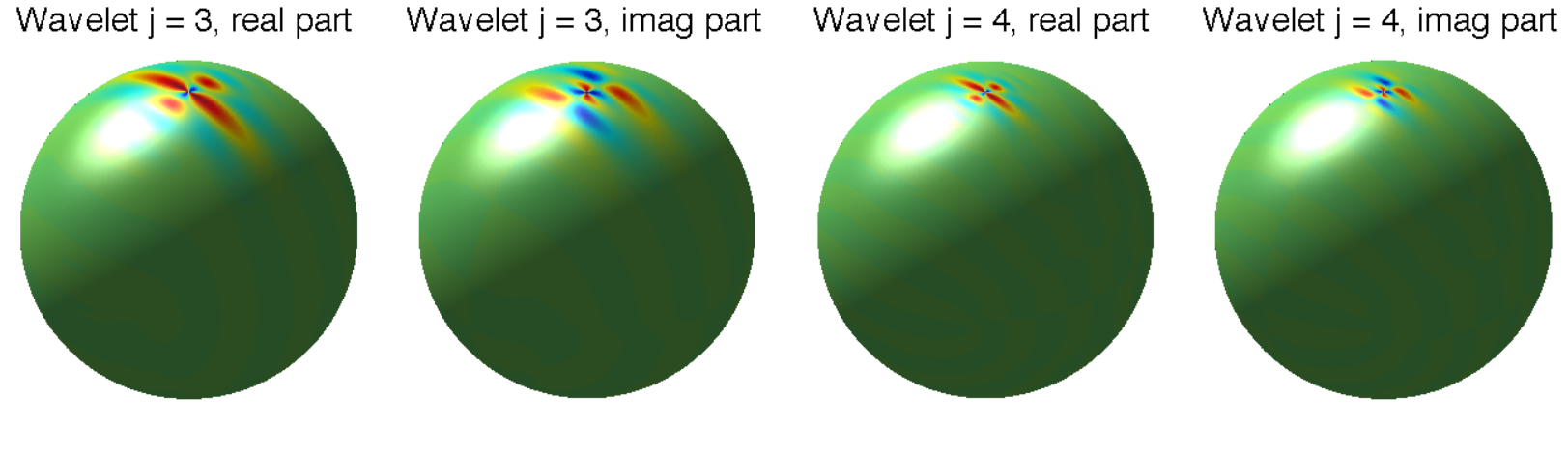}\end{minipage}}
\caption{Spin 0 and 2 wavelets constructed with angular band-limit $L=512$, azimuthal band-limit $N=15$, and tiling parameters $B=2$ and $J_{\rm min}=2$. }
\label{fig:wavs}
\end{figure}

We design a directional scale-discretised wavelet transform to analyse the directional features of signals of arbitrary spin on the sphere. Following Refs.~\cite{wiaux:2007:sdw, leistedt:s2let_axisym, mcewen:2013:waveletsxv, mcewen:2014:sccc21}, the spin-discretised wavelets ${}_s\Psi^j$ on the sphere are defined in harmonic space as ${}_s\Psi^j_{\ell m} \equiv \kappa^j(\ell) \xi_{\ell m}$, where $\kappa^j(\ell)$ characterises the angular localisation of the wavelets (for the $j$th scale), while $\xi_{\ell m}$ controls their directionality. Fig.~\ref{fig:wavs} shows examples of wavelets obtained with this construction. 

The wavelet transform of a spin signal ${}_sf$ on the sphere is given by the directional convolution with the wavelets. Provided the wavelets satisfy an admissibility property, the original signal can be synthesised exactly from its wavelet coefficients. More details about this transform can be found in Ref.~\cite{mcewen:2014:sccc21}.

\newpage 
\section{Application to the CMB}

The polarisation of the CMB is currently an intense avenue of research, since it may reveal signatures of primordial gravitational waves and a glimpse on the initial conditions of the universe. However, the physical quantities necessary for these investigations -- namely the E- and B-modes of the polarisation, or gradient and curl modes -- are not directly available. They are obtained by reducing and transforming frame-dependent observables: the Q and U ``local" polarisation measured on the sky. This Q-U to E-B transformation is unambiguous when dealing with data covering the entire sky. However, when CMB observations cover fractions of the sky only, E-B reconstruction is imperfect near the boundaries of the observation mask, causing leakage and potential biases in the recovered E- and B-mode maps. This issue is traditionally addressed by smoothing and extending the mask to remove boundary regions where the leakage is important. We develop a method exploiting the novel directional spin wavelet transform to decrease the leakage near the mask boundaries and obtain a more accurate E-B reconstruction. 

First, we compute the wavelet coefficients of the observable $Q+iU$ using our \emph{spin} wavelet transform. Second, we deal with the partial sky coverage by masking the data in spin wavelet space. Third, we apply to the real and imaginary parts of these wavelet coefficients a {\it scalar} inverse wavelet transform. It can be shown that this yields estimates of the E and B signals provided the wavelets of this inverse transform are spin-lowered versions of those used in the first transform Ref.~\cite{mcewen:2014:sccc21}.  This new method to estimate the E- and B-mode contributions from Q and U observations can significantly reduce the E-B leakage by exploiting improved masking in spin wavelet space. 

We anticipate our novel spin wavelet transform will be of general use in analysing CMB polarisation data, beyond this first application to E-B separation. A more detailed description of spin scale-discretised wavelets, fast algorithms, and a rigorous evaluation of their performance will be given in a series of forthcoming articles.

\bibliographystyle{IEEEtran}

\bibliography{bib_journal_names_short,bib_myname,bib}

\begin{thebibliography}{1}
\providecommand{\url}[1]{#1}
\csname url@samestyle\endcsname
\providecommand{\newblock}{\relax}
\providecommand{\bibinfo}[2]{#2}
\providecommand{\BIBentrySTDinterwordspacing}{\spaceskip=0pt\relax}
\providecommand{\BIBentryALTinterwordstretchfactor}{4}
\providecommand{\BIBentryALTinterwordspacing}{\spaceskip=\fontdimen2\font plus
\BIBentryALTinterwordstretchfactor\fontdimen3\font minus
  \fontdimen4\font\relax}
\providecommand{\BIBforeignlanguage}[2]{{%
\expandafter\ifx\csname l@#1\endcsname\relax
\typeout{** WARNING: IEEEtran.bst: No hyphenation pattern has been}%
\typeout{** loaded for the language `#1'. Using the pattern for}%
\typeout{** the default language instead.}%
\else
\language=\csname l@#1\endcsname
\fi
#2}}
\providecommand{\BIBdecl}{\relax}
\BIBdecl

\bibitem{wiaux:2007:sdw}
Y.~Wiaux, J.~D. McEwen, P.~Vandergheynst, and O.~Blanc, ``Exact reconstruction
  with directional wavelets on the sphere,'' \emph{MNRAS}, vol. 388, no.~2, pp.
  770--788, 2008.

\bibitem{leistedt:s2let_axisym}
B.~Leistedt, J.~D. McEwen, P.~Vandergheynst, and Y.~Wiaux, ``{S2LET}: A code to
  perform fast wavelet analysis on the sphere,'' \emph{A\&A}, vol. 558, no.
  A128, pp. 1--9, 2013.

\bibitem{mcewen:2013:waveletsxv}
J.~D. McEwen, P.~Vandergheynst, and Y.~Wiaux, ``On the computation of
  directional scale-discretized wavelet transforms on the sphere,'' in
  \emph{SPIE Wavelets and Sparsity XV}, 2013.

\bibitem{mcewen:2014:sccc21}
J.~D. McEwen, M.~{B\"{u}ttner}, B.~Leistedt, H.~V. Peiris, P.~Vandergheynst,
  and Y.~Wiaux, ``{On spin scale-discretised wavelets on the sphere for the
  analysis of CMB polarisation},'' in \emph{Proceedings IAU Symposium 306},
  2014, pp. 119--126.

\end{thebibliography}

\end{document}